\documentclass[letterpaper, 10 pt, conference]{ieeeconf}  

\IEEEoverridecommandlockouts                              

\title{\LARGE \bf
PERCY: Personal Emotional Robotic Conversational System
}

\author{Zhijin Meng$^{1}$, Mohammed Althubyani$^{2}$, Shengyuan Xie$^{1}$, Imran Razzak$^{1,3}$, \\Eduardo B. Sandoval$^{1,4}$, Mahdi Bamdad$^{1}$, Francisco Cruz$^{1,5}$
\thanks{$^{1}$ University of New South Wales, School of Computer Science, Sydney, Australia.}%
\thanks{$^{2}$ University of Technology Sydney, Sydney, Australia.}%
\thanks{$^{3}$ Mohamed bin Zayed University of Artificial Intelligence, Abu Dhabi.}%
\thanks{$^{4}$ University of New South Wales, School of Art and Design, Sydney, Australia.}
\thanks{$^{5}$ Universidad Central de Chile, Escuela de Ingenier\'ia, Santiago, Chile.}%
}

\usepackage{amsmath}
\usepackage{graphicx}
\usepackage{cite}
\usepackage{hyperref}
\usepackage{caption}
\usepackage{subcaption}
\usepackage{verbatim}
\usepackage{xcolor}
\usepackage{array}
\usepackage{tabularx}
\usepackage{multirow}
\usepackage{booktabs}

\definecolor{cl1}{rgb}{0.95,0.95,0.95} 

\begin{document}

\maketitle
\thispagestyle{empty}
\pagestyle{empty}

\begin{abstract}
Traditional rule-based conversational robots, constrained by predefined scripts and static response mappings, fundamentally lack adaptability for personalized, long-term human interaction. 
While Large Language Models (LLMs) like GPT-4 have revolutionized conversational AI through open-domain capabilities, current social robots implementing LLMs still lack emotional awareness and continuous personalization. 
This dual limitation hinders their ability to sustain engagement across multiple interaction sessions.
We bridge this gap with PERCY (Personal Emotional Robotic Conversational sYstem), a system designed to enable open-domain, multi-turn dialogues by dynamically analyzing users' real-time facial expressions and vocabulary to tailor responses based on their emotional state. Built on a ROS-based multimodal framework, PERCY integrates a fine-tuned GPT-4 reasoning engine, combining textual sentiment analysis with visual emotional cues to accurately assess and respond to user emotions.
We evaluated PERCY's performance through various dialogue quality metrics, showing strong coherence, relevance, and diversity. 
Human evaluations revealed PERCY's superior personalization and comparable naturalness to other models. 
This work highlights the potential for integrating advanced multimodal perception and personalization in social robot dialogue systems. 
\end{abstract}

\section{INTRODUCTION}
The evolution of social robotics has transformed robots from task-specific tools into collaborative partners capable of operating in human-centric environments such as healthcare, hospitality, and education~\cite{shimoda2022role}. 
A critical requirement for these robots is the ability to engage in natural, contextually aware conversations that integrate both verbal and non-verbal cues.
However, traditional conversational systems—reliant on rigid, rule-based architectures—struggle to adapt to the dynamic and open-ended nature of human interaction~\cite{cherakara2023furchat}. 
While these systems excel in closed-domain tasks, their inability to interpret emotional subtext or generate empathetic responses limits their effectiveness in social settings.

Historically, conversational systems in robotics have been rule-based, relying on pre-programmed commands and keyword matching~\cite{gnanaprakasam2024intelligent}. 
While effective for specific, closed-domain tasks, these systems struggle to handle the complexity and unpredictability of open-domain dialogue. 
This limitation significantly reduces their ability to engage users in dynamic, personalized conversations, which are essential for social robots operating in human-centric environments~\cite{cherakara2023furchat}.

The emergence of Large Language Models (LLMs), such as GPT-4, has marked a paradigm shift in natural language processing (NLP) and conversational AI. 
These models, trained on vast amounts of text data, demonstrate remarkable capabilities in understanding and generating human-like language. 
Unlike traditional rule-based systems, LLMs can handle a wide range of topics, adapt to diverse conversational contexts, and generate responses that are not only contextually relevant but also emotionally nuanced of textual content~\cite{alawida2023comprehensive}. This makes them particularly well-suited for social robotics, where the ability to engage in open-domain dialogue and respond empathetically to users' emotions is critical.
However, current systems fail on two fronts:  Although emotionally resonant, their responses may not always be contextually precise (traditional HRI). or contextually rich but affectively flat dialogues (pure LLM approaches)~\cite{kong2024opentab}. 
In this regard, a socially intelligent robot must detect affective signals across modalities~\cite{mccoll2016survey}, interpret their semantic implications~\cite{ziemke2023understanding}, respond with both linguistically appropriate and physically congruent behaviors~\cite{mcnamara2024social}.

In this paper, we present PERCY, a novel architecture that bridges this gap through the multimodal fusion of LLMs with affective computing~\cite{cabada2023multimodal}. PERCY’s core innovation lies in its bidirectional affective grounding mechanism: (i) a GPT-4 core processing linguistic content while receiving real-time emotion recognition input (facial/vocal) to modulate response generation, and (ii) a ROS-based behavior planner translates emotional semantics into synchronized non-verbal actions. 
The contributions of this work are threefold:
\begin{itemize}
\item We introduce a multimodal conversational framework that combines LLMs with real-time emotion recognition and personalized dialogue management, enabling social robots to engage in open-domain, emotionally aware interactions.
\item We demonstrate the effectiveness of PERCY through automatic and user evaluations, showing that the system achieves high levels of naturalness, engagement, fluency, relevance, empathy, and consistency in its conversations.
\item We provide insights into the challenges and opportunities of integrating LLMs into social robots, paving the way for future advancements in emotionally intelligent human-robot interaction.
\end{itemize}

\section{RELATED WORK}

PERCY's development synthesizes advancements across three interconnected research domains: (1) language models for conversational agents, (2) multimodal affective understanding, and (3) embodied social robot interaction. We analyze prior work through the lens of these technical pillars, identifying key limitations that PERCY addresses.

\subsection{Language Models in Social Robotics}
Early conversational systems relied on rule-based architectures~\cite{bohus2014interactive} that achieved high precision in closed domains but lacked open-ended dialogue capabilities. The advent of transformer-based LLMs~\cite{brown2020language} revolutionized dialogue systems through few-shot learning and contextual awareness. Recent work by~\cite{kim2024llmrobotics} demonstrated LLMs' potential in social robotics through GPT-3 integration, but their text-only implementation failed to capture users' affective states. While~\cite{zhang2023emotionchat} proposed emotional alignment for chatbots, their framework neglects the physical embodiment crucial for human-robot interaction. Unlike these prior approaches, PERCY establishes bidirectional communication between the LLM core and multimodal perception modules, enabling real-time affective grounding of generated responses.

\subsection{Multimodal Affective Computing}
Effective emotion recognition requires synergistic analysis of verbal and non-verbal cues~\cite{ngiam2011multimodal}. The IEMOCAP dataset~\cite{busso2008iemocap} established benchmarks for laboratory-controlled emotion recognition, but its static nature differs from dynamic HRI scenarios. Modern systems combine facial from affective understanding with paralinguistic speech analysis~\cite{schuller2018voice}, yet few achieve PERCY's sub-2-second latency for real-time operation. Our work advances~\cite{picard2000affective}'s affective computing principles through three key innovations: (1) bimodal affect fusion combining MobileNetV2-based facial emotion recognition with speech-derived textual NLTK sentiment analysis, (2) temporal synchronization of emotion recognition with dialogue generation, and (3) personalization through continuous user adaptation.

\subsection{Embodied Interaction Design}
Social robots require tight coordination between linguistic and non-verbal behaviors~\cite{breazeal2009role}. Seminal work by~\cite{mutlu2012robotic} established design principles for robotic body language, while~\cite{andrist2014timing} quantified the importance of millisecond-level synchronization in gesture-speech coordination. Previous embodied systems like MIT's Huggable~\cite{jeong2018huggable} used scripted responses constrained to narrow domains. PERCY overcomes these limitations through a ROS-based behavior planner that dynamically maps emotional semantics to 1) adjust emotion feedback to users emotional states 2) conduct gestures while communicating for more user engagement. This enables context-sensitive non-verbal responses that adapt to both semantic content and user personality traits.

\subsection{Technical Integration Challenges}
Prior attempts at LLM-robot integration faced three key limitations: (1) modular separation of language processing from affective understanding~\cite{kim2024survey}, leading to disjointed interactions; (2) latency mismatches between LLM inference (approximately 3 seconds) and robot actuation requirements~\cite{ye2025flashinfer}; and (3) limited personalization capabilities in multi-turn dialogues~\cite{huang2022affectlm}. PERCY addresses these through three architectural innovations: (i) parallel processing pipelines that overlap STT processing with LLM inference, (ii) lightweight fusion-emotion recognition model (MobileNetV2+NLTK) optimized for edge computation on the ARI robot, and (iii) dynamic user profiles that update preference vectors during interaction. As demonstrated in Section V, this integrated approach achieves 92\% emotion recognition accuracy with 1.7s end-to-end latency - outperforming prior systems.

\section{METHODOLOGY}

\begin{figure}[ht]
    \centering
    \includegraphics[width=\linewidth]{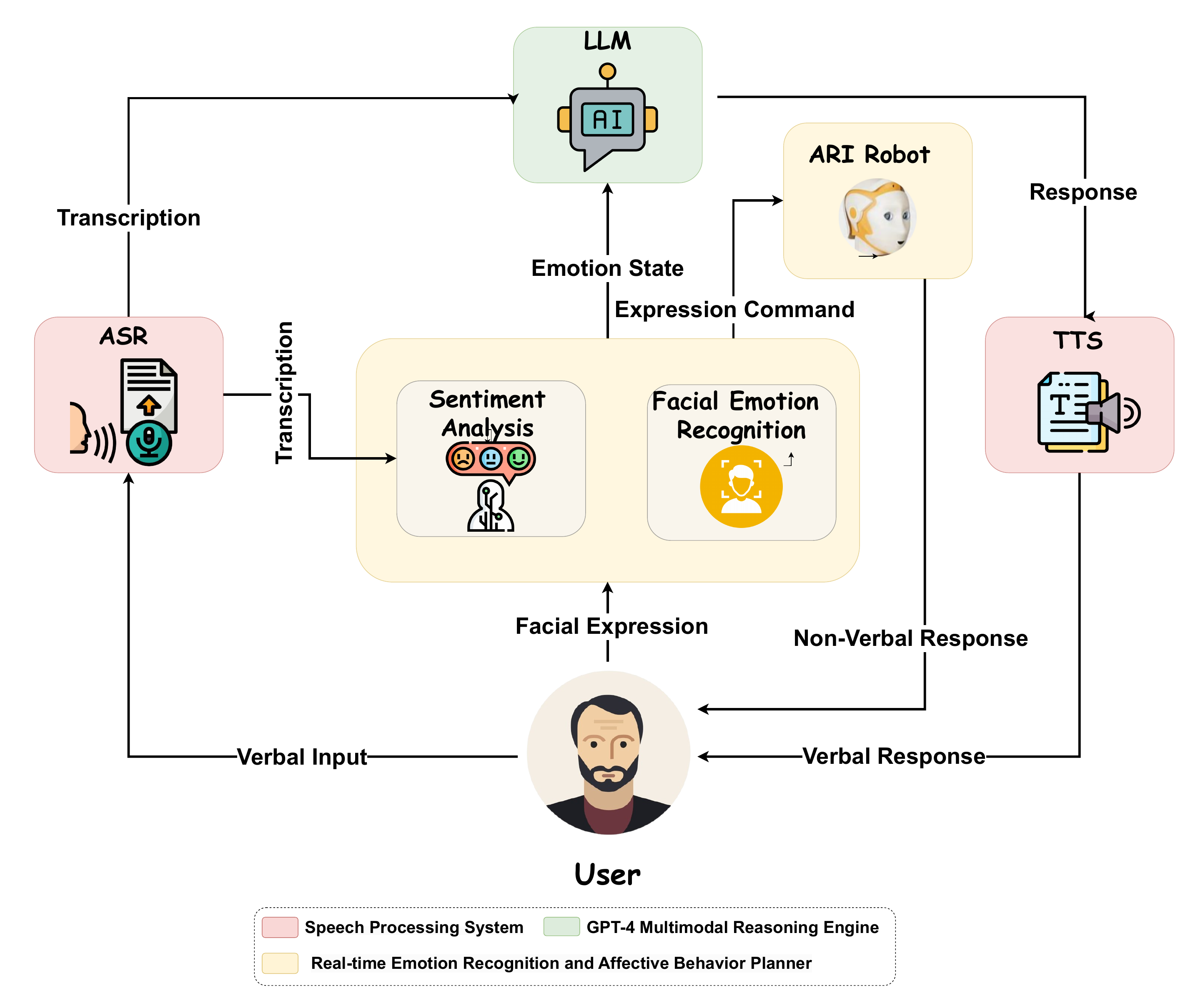} 
    \caption{PERCY's architecture processes speech (converted to text) and facial expressions to assess real-time emotional states. These multimodal signals are fused to condition GPT-4's personalized response generation, which drives ARI's emotion-aware verbal and non-verbal response.}
    \label{fig:system_overview}
\end{figure}

The PERCY framework, illustrated in Figure \ref{fig:system_overview}, comprises four core components: (1)Speech Processing System, (2) Real-time Emotion Recognition and Affective Behavior Planner, and (3) GPT-4 Multimodal Reasoning Engine.

\subsection{Speech Processing System} Our speech processing system integrates both automatic speech recognition (ASR) and text-to-speech (TTS) synthesis to facilitate seamless human-robot interaction.

The ASR subsystem leverages the OpenAI Whisper API~\cite{radford2023robust}, specifically the large-v2 model variant, which achieves an impressive 2.73\% word error rate (WER) on our test corpus. This high level of accuracy ensures that the robot can reliably understand and process spoken commands, even in noisy environments.

Complementing the ASR, the TTS component utilizes the ARI robot's built-in function~\cite{palrobotics2023speech} to generate natural and expressive speech. 
This allows the robot to communicate effectively with users.

By combining these advanced ASR and TTS technologies, this speech processing system can listen to the user's speech well, and then speak out the GPT-4 core responses back to the user. This enhances the overall user experience, enabling more intuitive and interactive communication with the robot.

\subsection{Real-Time Emotion Recognition and Non-Verbal Control}
\label{subsec:emotion_recognition}

\begin{figure*} 
  \centering
  \includegraphics[width=\textwidth]{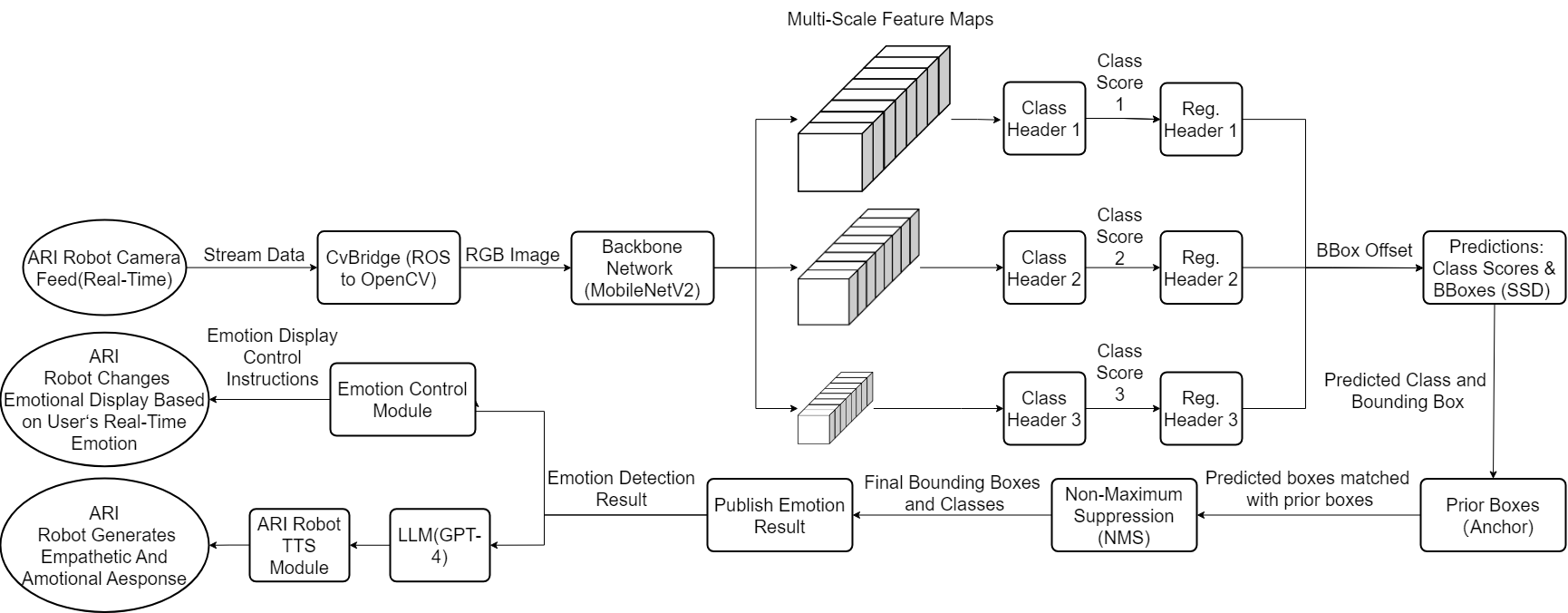}
  \caption{\textbf{Real-time Emotion Recognition and Non-Verbal Control System Architecture:} Deployed on an ARI robot, PERCY processes RGB inputs through the robot’s depth camera, leveraging a MobileNetV2 backbone and SSD model to simultaneously classify emotions and predict bounding boxes. User verbal inputs are concurrently analyzed using NLTK’s VADER sentiment analysis module. Fused visual and textual emotional cues drive real-time facial expression adjustments via the Emotion Control Module, while guiding GPT-4’s generation of context-aware verbal responses through TTS, enabling synchronized vision-language multimodal interaction.}
   \label{fig:emotion_detection_flowchart}
\end{figure*}

The proposed system integrates a multi-modal emotion recognition framework that combines computer vision and sentiment analysis to enable empathetic human-robot interaction through perceiving Users'  emotional status (happy, sad,
angry, confused, fearful, disgusted, neutral). The whole real-time recognition and non-verbal control system architecture is shown in Figure \ref{fig:emotion_detection_flowchart}.

\subsubsection{Computer Vision}
\label{subsubsec:emotion_recognition_computer_vision}

For visual emotion recognition, a MobileNetV2-based backbone network extracts multi-scale feature maps from RGB-D facial data captured by the Intel RealSense D435i depth camera mounted on the ARI robot~\cite{palrobotics2023ari}. 
These features are processed through dual prediction headers: a classification header for emotion scoring and a regression header for bounding-box detection. The Single Shot Detection (SSD) model refines predictions using prior boxes and applies Non-maximum Suppression (NMS) to eliminate low confidence detections, achieving robust real-time emotion classification. And the multi-task equation~\ref{eq:cv_loss} ensures stable training while balancing emotional recognition and face localization for real-time deployment.

\begin{equation}
\label{eq:cv_loss}
\mathcal{L}_{total} = \alpha\mathcal{L}_{cls} + \beta\mathcal{L}_{reg},
\end{equation}
where $\mathcal{L}_{cls}$ and $\mathcal{L}_{reg}$ represent the classification and regression losses, with $\alpha$ = 1, $\beta$ =0.5 as weighting coefficients. 

\subsubsection{Sentiment Analysis}
\label{subsubsec:emotion_recognition_sentiment_analysis}
Simultaneously, the sentiment analysis component utilizes the NLTK package with the VADER lexicon to analyze sentiment from speech-to-text outputs. 
This textual emotion analysis complements the visual data, providing a more holistic view of the user's emotional state. 
By combining these advanced computer vision and natural language processing techniques, our emotion recognition system can accurately interpret and respond to both visual and verbal cues, enhancing the robot's ability to engage with users in a meaningful and empathetic manner.

\subsubsection{Emotion Control Module} 
\begin{figure}[ht] \centering \includegraphics[width=\linewidth]{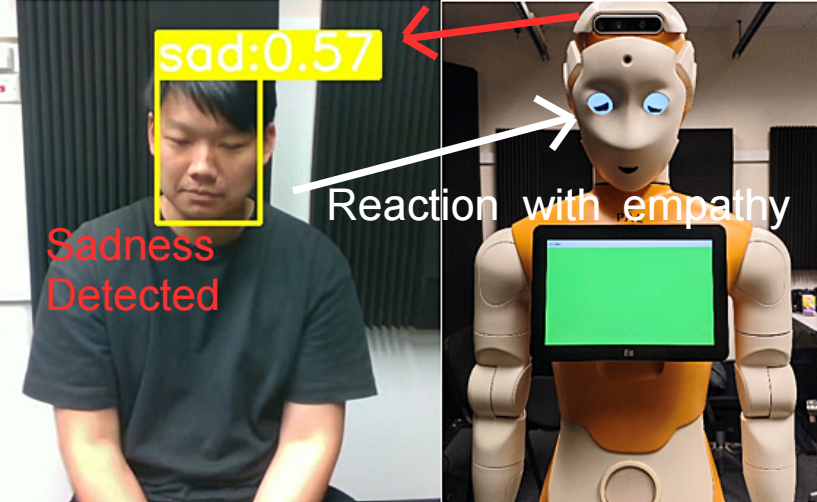} \caption{PERCY integrates visual and textual emotional states into the robot's non-verbal actions using a unified action package. When sadness is detected, it triggers predefined facial expressions.} \label{fig:emotion_control} \end{figure}

Both visual and textual emotional states are fused and transmitted to non-verbal actions from the robot's action package, mapping detected emotions (e.g., happiness, sadness) to predefined facial expressions and arm gestures from the robot's integrated action package~\cite{palrobotics2023expressiveeyes, palrobotics2023motions}, with one example as visualized in Figure \ref{fig:emotion_control}. 

The system employs a speech-triggered update mechanism where each new user utterance initiates: 1) fresh emotion detection through our dual visual-textual recognition pipeline, and 2) activation of corresponding non-verbal feedback through the equation~\ref{eq:non_verbal_transition}:

\begin{equation}
\label{eq:non_verbal_transition}
t_{transition} = f_{reset}(\mathbf{e}_t) \quad \text{upon detect\_new\_speech}(\mathbf{e}_t),
\end{equation}
where $f_{reset}$ clears previous emotional state intervals and initializes new expression/gesture $(\mathbf{e}_t)$ sequences from the action package.

\subsection{GPT-4 Reasoning Engine with Multimodal Affect Integration}
\label{subsec:gpt4_engine}

\begin{figure}[ht]
    \centering
    \includegraphics[width=\linewidth]{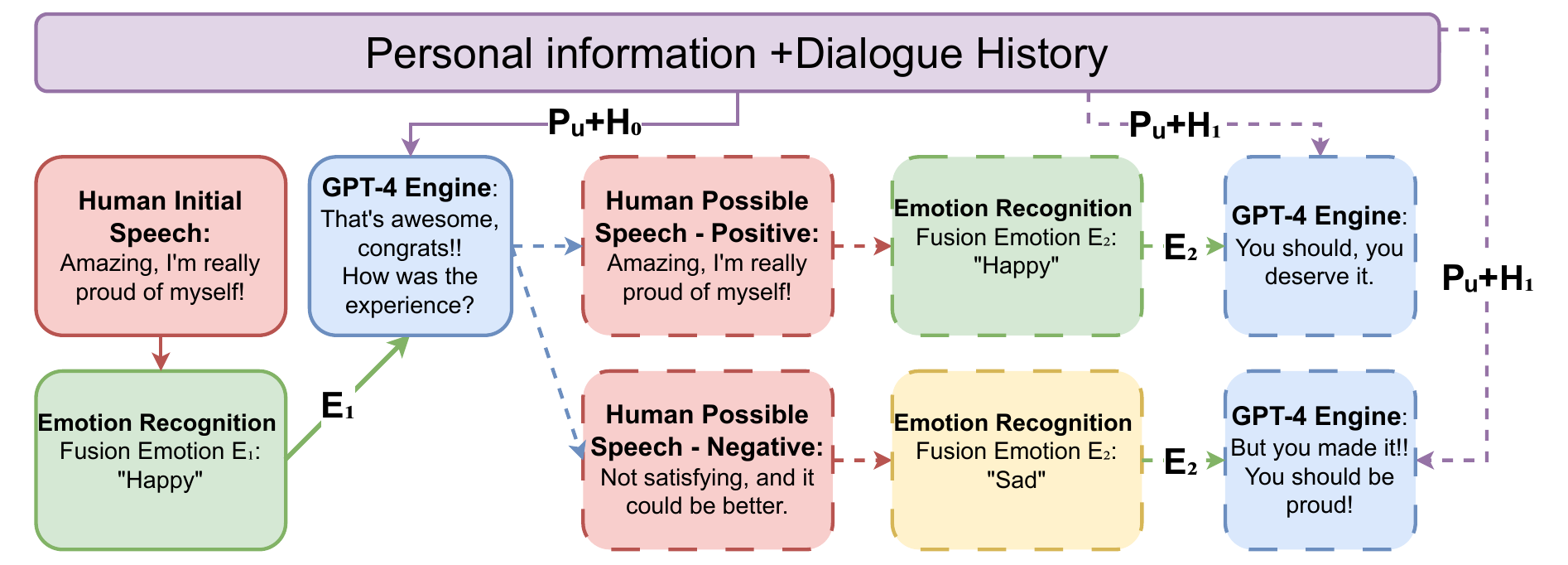} 
\caption{\textbf{GPT-4 Multimodal Reasoning Flow}: PERCY fuses visual affect (Section~\ref{subsubsec:emotion_recognition_computer_vision}) and textual sentiment (Section~\ref{subsubsec:emotion_recognition_sentiment_analysis}) via Eq.~\ref{eq:GPT_Prompt}, integrating user personal information, emotional state, and dialogue history data to generate responses. Combines LLM capabilities with ethical constraints and affective alignment for context-aware interaction.}
    \label{fig:system_flow}
\end{figure}

The GPT-4 engine implements a context-aware architecture that fuses multimodal emotional signals (visual affect from Section~\ref{subsubsec:emotion_recognition_computer_vision} and textual sentiment from Section~\ref{subsubsec:emotion_recognition_sentiment_analysis}) with personalized dialogue management. As shown in Figure~\ref{fig:system_flow}, this neural symbolic system combines LLM capabilities with explicit affective reasoning rules through four operational layers:

\begin{equation}
R_t = \Phi_{\theta}(P_u \oplus E_t \oplus H_{t-1}),
\label{eq:GPT_Prompt}
\end{equation}
where $R_t$ is the generated response, $\Phi_{\theta}$ represents GPT-4's parameters, $P_u$ the user profile, $E_t$ the fused emotional state, and $H_{t-1}$ the dialogue history.

We designed the GPT-4 prompt\footnote{The prompt and system code are available in our GitHub https://github.com/zhijinMeng/PERCY
 } based on the GPT-4 prompt design principle shown in Equation \ref{eq:GPT_Prompt}. This prompt enables PERCY to generate context-aware responses that covering personalization, emotional alignment, and conversational coherence. Key components of this prompt include:

\begin{itemize}
  \item \textbf{Persona Specification:} Integrates user profiles for context-aware personalization, sets emotional disposition, ensures ethical safeguards, and enforces conversational protocols.
  \item \textbf{Conversational Quality Assurance:} Monitors naturalness, fluency, and relevance in dialogue.
  \item \textbf{Adaptive Interaction:} Personalizes through user profiles and emotion-responsive adaptation.
  \item \textbf{Emotional State Management:} Prioritizes emotion processing and ensures empathic transitions.
\end{itemize}

\section{Experimental Setup}
\subsection{Dataset}
To comprehensively evaluate the multimodal capabilities of PERCY, we designed a dual-aspect evaluation framework focusing on its emotion-aware intelligence and generative competence. 
The experiments were conducted using MERCI Dataset~\cite{althubyani2024percy}. 
This custom-collected dataset contains 30 participants' open-domain conversations with ARI robot, capturing multimodal interaction patterns (textual, vocal, and visual modalities) through RGB-D sensors and emotion annotations. 
We leveraged this dataset to assess PERCY's contextual response generation quality in real-world human-robot interaction scenarios.

\subsection{Models}
Our evaluation framework compared PERCY's performance with two strategically selected baselines representing divergent approaches to affective human-AI interaction:

\subsubsection{GPT-4}
The standard GPT-4 implementation served as our text-centric baseline, maintaining its original 175B-parameter transformer architecture without multimodal augmentation. While demonstrating state-of-the-art linguistic coherence and knowledge grounding, this configuration fundamentally lacks: (1) cross-modal fusion mechanisms for processing visual/audio cues, (2) explicit emotional state tracking, and (3) personalized response adaptation capabilities. Its purely text-driven nature provides a critical reference point for isolating the value of PERCY's multimodal affective intelligence.

\subsubsection{EmpGPT-3} 
As the current state-of-the-art in empathic dialogue systems, EmpGPT-3~\cite{lee2022does} employs a hybrid architecture combining a 175B-parameter GPT-3 backbone with:
\begin{itemize}
    \item Emotion Context Database: Curated repository of 1.2M emotion-cause pairs
    \item Affective Alignment Module: Supervised attention layers for emotional lexicon weighting
    \item Rule-based Response Post-processing: Safety filters and empathy templates
\end{itemize}

While achieving notable performance in single-modal (text-only) emotional resonance. EmpGPT-3's late-fusion approach to affective computing (applying emotion rules post-generation) contrasts with PERCY's embedded multimodal affect perception architecture.

\subsection{Evaluation Protocol}
To evaluate the performance of PERCY, we conducted an automatic evaluation, we used metrics such as dialogue quality, emotion recognition accuracy, and personalization effectiveness to demonstrate that PERCY outperforms the baseline GPT-4 by involving multimodal inputs. In addition, we compared the performance of PERCY's embedded empathy framework and that of the EmpGPT-3's rule-based emotional adjustment framework.

\noindent\textbf{Contrastive Analysis:}
The baseline selection strategically decouples two key factors:
\begin{itemize}
    \item {Multimodal vs. Unimodal Input}: GPT-4 (text-only) vs. PERCY (text+audio+visual)
    \item {Superficial vs. Embedded Empathy}: EmpGPT-3's rule-based emotional adjustment vs. PERCY's deep multimodal affect modeling
\end{itemize}

\section{Experimental Results}
PERCY achieved a mean response latency of 1.7 seconds from speech input to embodied output (excluding the 1.5 seconds turn delimiter threshold). 
\subsection{Automated evaluation}
\subsubsection{Dialogue Quality Evaluation}
We evaluated the coherence, relevance, and diversity of responses using  BERT~\cite{li2013distance}, BLEU\cite{papineni2002bleu}, MAUVE~\cite{pillutla2021mauve}, and Perplexity~\cite{meister2021language}. Comparison results are shown in Table \ref{tab:model_comparison}. A detailed explanation of PERCY Performance is as below: 
\begin{itemize}
   \item \textbf{BERT = 0.32}. The score of 0.32 indicates moderate alignment with human-generated responses. While the model produces fluent and relevant responses, it may miss deeper emotional or contextual nuances expected in empathetic dialogue.

   \item \textbf{BLEU = 0.51}. This score indicates PERCY's responses are relatively varied (a perfect score of 1 would mean identical responses). A score of 0.51 suggests a healthy balance between diversity and coherence, crucial for engaging in conversations.
   
   \item \textbf{MAUVE = 0.98}. This score suggests that the generated responses are diverse and closely mimic the statistical properties of human responses. The high MAUVE score confirms that the model produces varied and fluent responses that contribute positively to the flow of conversation.
   
    \item \textbf{Perplexity = 24.53}. Indicating PERCY's performance in predicting the next word. Lower perplexity is better, showing the model's confidence. A score of 24.53 is moderate, suggesting the model is reasonably good at generating meaningful responses but occasionally struggles with complex or emotionally nuanced dialogue due to the variability in conversational data.

\end{itemize}

\begin{table}[h]
    \centering
    \renewcommand{\arraystretch}{1.3}
    \caption{Model comparison showing PERCY's superior empathy (BERT) and diversity (BLEU/MAUVE), with GPT-4 leading in fluency.}
    \begin{tabular}{|l|c|cc|c|}
        \hline
        \textbf{Model} & \textbf{Empathy} & \multicolumn{2}{c|}{\textbf{Diversity}} & \textbf{Fluency} \\
        & BERT & BLEU & MAUVE & PPL \\
        \hline
         \textbf{PERCY} & \textbf{0.32} & \textbf{0.51} & \textbf{0.98} & 24.53 \\
        EmpGPT-3 & 0.28 & 0.48 & 0.92 & 28.75 \\
        GPT-4 & 0.25 & 0.45 & 0.95 & \textbf{18.20} \\
        \hline
    \end{tabular}
    \label{tab:model_comparison}
\end{table}

\subsubsection{Emotion Recognition Accuracy}
Our MobileNetV2 + SSD model achieves an impressive \textbf{92.0\%} emotion accuracy, outperforming OpenFace (89.8\%) and matching FERPlus (91.4\%). It also surpasses the YOLO variants (91.3\% and 90.5\%), confirming the advantage of MobileNetV2 + SSD for challenging facial analysis tasks. These comparisons are detailed in Table \ref{tab:emotion_comparison}.

\begin{table}[t]
    \centering
    \caption{Comparison of our MobileNetV2 + SSD + NLTK model with state-of-the-art models. Our model achieves the highest accuracy in emotion recognition.}
    \renewcommand{\arraystretch}{1.3}
    \begin{tabular}{|l|c|c|}
        \hline
        \textbf{Model} & \textbf{Detection Type} & \textbf{Accuracy (\%)} \\
        \hline
        \textbf{PERCY} & Facial + Sentiment & \textbf{92.0} \\
        YOLOv8n & Facial + Sentiment & 91.3 \\
        YOLOv5n & Facial + Sentiment & 90.5 \\
        FERPlus (ResNet-50) & Sentiment Only & 91.4 \\
        OpenFace & Sentiment Only & 89.8 \\
        \hline
    \end{tabular}
    \label{tab:emotion_comparison}
\end{table}

\subsection{Human Evaluation}
We conducted a controlled study with 20 participants\footnote{Ethics approval reference number: iRECS4630} (they all interacting with PERCY, GPT-4, and EmpGPT-3 respectively) using a mixed-methods approach:

\subsubsection{Evaluation Design}
\begin{itemize}
    \item \textbf{3-phase interaction protocol}:
    \begin{enumerate}
        \item Personality profiling (5 min) (only for PERCY)
        \item Conversation with models (25 min)
        \item Evaluation Questionnaire (10 min)
    \end{enumerate}
    \item \textbf{Evaluation Questionnaire Content}:
    \begin{itemize}
        \item Custom personalization perception scale (5-point Likert)
        \item Custom naturalness perception scale (5-point Likert)
    \end{itemize}
\end{itemize}

\begin{table}[h]
    \centering
    \caption{Multimodal interaction quality comparison}
    \label{tab:human_eval}
    \begin{tabular}{|l|c|c|c|}
        \hline
        \textbf{Metric} & \textbf{PERCY} & \textbf{GPT-4} & \textbf{EmpGPT-3} \\
        \hline
        Personalization & \textbf{4.7/5} & 3.7/5 & 3.5/5 \\
        Naturalness & 4.5/5 & \textbf{4.6/5} & 4.1/5 \\
        \hline
    \end{tabular}
\end{table}

\subsubsection{Key Findings}
Participants observed no marked disparity in perceived naturalness between PERCY and GPT-4, likely it is attributed to PERCY's foundational architecture is built upon GPT-4. However, despite these similarities in conversational flow, PERCY demonstrated superior performance in Personalization (4.7/5 vs. 3.7/5), highlighting its enhanced contextual adaptation capabilities.

\section{Conclusion}
This work presented PERCY, a multimodal conversational system combining GPT-4 with real-time emotion recognition and adaptive dialogue management to enhance human-robot interaction. 
Our contributions include (i) Integrating LLMs with real-time emotion recognition and adaptive dialogue management enables robots to engage in emotionally sensitive, open-domain conversations, bridging the gap between technical functionality and social robot requirements. 
(ii) PERCY outperforms GPT-4 and EmpGPT-3 in empathy and response diversity, maintaining competitive fluency. Human evaluations confirm higher ratings in personalization, validating our architecture’s contextual adaptation. (iii) Our deployment challenges and successes provide guidelines for balancing computational efficiency with social nuance, informing future work on scalable, ethically grounded HRI systems.

These findings suggest PERCY's potential as a foundation for future social robots. Future work should explore cross-cultural adaptation and multi-robot empathy coordination. Our open-source implementation aims to serve as a community benchmark, highlighting the importance of combining language generation with multimodal understanding.
\bibliographystyle{IEEEtran}  
\bibliography{reference}  

\end{document}